\documentstyle[twoside,fleqn,espcrc2]{article}
\begin{document}

\def\dsl{\raise.15ex\hbox{/}\kern-.57em\partial}
\def\NPB#1#2#3{{\sl Nucl. Phys.} \underbar{B#1} (#2) #3}
\def\PLB#1#2#3{{\sl Phys. Lett.} \underbar{#1B} (#2) #3}
\def\PRL#1#2#3{{\sl Phys. Rev. Lett.} \underbar{#1} (#2) #3}
\def\PRD#1#2#3{{\sl Phys. Rev.} \underbar{D#1} (#2) #3}
\def\CMP#1#2#3{{\sl Comm. Math. Phys.} \underbar{#1} (#2) #3}
\def\hepth#1{{hep-th/#1}}
\def\hepph#1{{hep-ph/#1}}
\def\del{\partial}
\def\bar{\overline}
\def\bC{{\bf C}}
\def\bZ{{\bf Z}}
\def\ie{{\it i.e.}}
\def\eg{{\it e.g.}}
\def\til{\widetilde}
\def\tA{{\til A}}
\def\tF{{\til F}}
\def\DD{{\cal D}}
\def\MM{{\cal M}}
\def\CO{{\cal O}}
\def\WW{{\cal W}}
\def\semi{;\hfil\break}
\def\tQ{{\til Q}}
\def\tq{{\til q}}
\def\vev#1{{\langle #1\rangle}}
\def\bea{\begin{eqnarray}}
\def\eea{\end{eqnarray}}
\def\nn{\nonumber}
\def\be{\begin{equation}}
\def\ee{\end{equation}}
\def\bc{\begin{center}}
\def\ec{\end{center}}

\input epsf.tex

\title{Aspects of N=1 String Dynamics \thanks{Lectures presented at
the 33rd Karpacz Winter School ``Duality: Strings and Fields'' (Poland,
February 1997).}}

\author{Shamit Kachru\address{Department of Physics,
        Rutgers University, Piscataway, NJ 08855}}

\begin{abstract}

We review several topics of interest in the study of 4d N=1 supersymmetric
compactifications of the heterotic string.  
After a brief introduction to the construction of such models, our focus
is on the novel physics which occurs at singularities in the moduli space
of vacua.  Among the phenomena we discuss are nonperturbative superpotentials,
dynamical generation of poles in various low-energy couplings, and phase 
transitions which change the net number of chiral generations.

\end{abstract}

\maketitle

\section{Introduction}

Much has been learned from string duality about superstring vacua with
extended supersymmetry (for reviews of various aspects, see e.g. \cite{revs}).
For instance: 

$\bullet$ Naively distinct moduli spaces are smoothly connected
(e.g. through 
conifold transitions in 4d N=2 type II vacua \cite{old,andy,gms})

$\bullet$ Dual descriptions allow one to compute nonperturbative
effects in one picture by doing tree-level computations in a dual theory
(e.g. in heterotic/type II duality in four dimensions with N=2 
supersymmetry \cite{kv,fhsv})

While we now have an excellent understanding of most phenomena which
occur in string vacua with extended supersymmetry, our understanding of
4d N=1 vacua is much less complete. 
There are several motivations for studying such vacua:

$\bullet$ Less supersymmetry allows richer phenomena.  For example,
in N=1 theories the classical moduli space can be
$\it lifted$ by quantum effects, i.e. supersymmetry can be dynamically
broken.  This is impossible in theories with extended supersymmetry. 

$\bullet$ 4d N=1 is the case which might be relevant to the real world.
N=1 allows chiral gauge representations, while for extended supersymmetric
theories one typically has only nonchiral representations.  In addition,
weak scale SUSY is one plausible explanation for the unnaturally large
ratio of the weak scale to the Planck scale, the so-called ``hierarchy
problem.''  Finally, the observed unification of the gauge couplings in
minimal supersymmetric extensions of the standard model leads one to
suspect that N=1 SUSY may be lurking around the corner at the weak scale.

$\bullet$ N=1 SUSY is much more ``generic'' than extended SUSY,
just in terms of number of vacua.  Such vacua should be studied for this
reason alone.

\section{N=1 Theories from String Compactification}

There are several different ways to get 4d N=1 theories out of 
string theory and its parent theories, M-theory and F-theory.
We describe a few methods of constructing N=1 theories below.
For most of these lectures, our focus will be on the heterotic 
$SO(32)$ and $E_8 \times E_8$ strings.

\subsection{Heterotic Theories}

One can start with either $E_8\times E_8$ or $SO(32)$ gauge group in
ten dimensions. 
Then the low energy effective
theory is 10d N=1 Super Yang-Mills, coupled to supergravity.
If we look for solutions of this theory preserving one four-dimensional
supersymmetry, we find \cite{chsw}:

$\bullet$ The compactification space $M$ must be a manifold of
$SU(3)$ holonomy, a 
Calabi-Yau manifold.  
The reduced holonomy implies that the first Chern class $c_{1}(M)$
vanishes.
Such manifolds are
complex Kahler manifolds and can be
equipped with a Ricci-flat metric for each choice of their 
complex structure and Kahler class $J$.  
The parameters involved in these choices yield scalar moduli in
the 4d N=1 effective field theory.  One obtains $h^{2,1}(M)$ such
moduli parametrizing the choice of complex structure, and
$h^{1,1}(M)$ giving the choice of Kahler class $J$.

$\bullet$ The gauge 
field strength $F$ must satsify the equations
$$F_{ij} = F_{\bar i \bar j} = 0, ~~g^{i \bar j}F_{i \bar j} = 0
$$
where 
$i$ and $\bar i$ label (anti)holomorphic coordinates on the complex
manifold $M$ and $g$ is the Ricci-flat Kahler metric on $M$.
These equations have a solution iff the gauge connection is a connection on a
$\it stable,~holomorphic$ vector bundle $V \rightarrow M$ \cite{duy}.  
Furthermore,
there is an integrability condition for the solution of the Kahler
Yang-Mills equations
$$\int_{M} J\wedge J \wedge c_{1}(V) ~=~0
$$
Also, the Bianchi identity for the three-form field strength $H$
requires
$$dH = Tr(R\wedge R) - Tr(F\wedge F) ~\rightarrow ~c_{2}(TM) ~=~c_{2}(V)
$$
Typically, the vector bundles $V$ come in multiparameter families 
parametrized by
elements of $H^{1}(M, End(V))$.

\centerline{\bf{EXAMPLES}}

There are relatively simple techniques for constructing many
examples of Calabi-Yau threefolds; for a deeper understanding, 
one should consult e.g. \cite{hubsch}.
The simplest examples involve constructing the Calabi-Yau
as a hypersurface or a complete intersection of hypersurfaces in
some $\it {weighted~projective~space}$. 

The projective space $CP^{N}$ is simply a space defined by
homogeneous complex coordinates $z_{1},\cdots,z_{N+1}$ subject
to the relation
$$(z_{1},\cdots,z_{N+1}) \sim (\lambda z_{1},\cdots,\lambda z_{N+1})$$
for any nonzero complex $\lambda$, and with the origin removed. 
Using the adjunction formula \cite{hubsch}, it is easy to see
that a hypersurface $M$ defined by an equation which is
homogeneous of degree
$N+1$ in the $z_{i}$
will have $c_{1}(M)=0$.  To get a threefold in this way, 
one needs to work in $CP^4$, and one obtains the
famous quintic in $CP^4$.  Similarly, any intersection of
$k$ hypersurfaces, with defining polynomials whose degrees sum
to $N+1$, will define a Calabi-Yau $N-k$ fold in $CP^N$.

A simple generalization of the complex projective space $CP^N$
is the weighted projective
space $WCP^{N}_{w_{1},\cdots,w_{N+1}}$ defined by homogeneous coordinates
$z_{1},\cdots,z_{N+1}$ subject to an identification
$$(z_{1},\cdots,z_{N+1}) \sim (\lambda^{w_{1}}z_{1},\cdots,\lambda^{w_{N+1}}
z_{N+1})$$
for any nonzero complex $\lambda$ (and again with the origin
removed).  
One calls a polynomial $F(z_i)$ homogeneous of degree $d$ in this weighted
projective space if
$$F(\lambda^{w_1}z_1,\cdots,\lambda^{w_{N+1}}z_{N+1}) \sim 
\lambda^{d}F(z_{1},\cdots,
z_{N+1})$$ 
It again follows from
the adjunction formula that a 
complete intersection of $k$ hypersurfaces whose
degrees add up to $\sum w_{i}$ will 
define a Calabi-Yau $N-k$ fold in the $WCP^N$. 
There are subtleties
in certain cases, involving singularities of the ambient $WCP^N$
which might be inherited by the Calabi-Yau; we will not discuss
these issues here, and refer the reader to \cite{hubsch}.

We have not mentioned
the construction of appropriate gauge bundles here.  Each Calabi-Yau
comes equipped with its tangent bundle, which serves as a suitable
choice.  
However, one should keep in mind that this choice, ``embedding the
spin connection in the gauge group,'' gives only a very small
class of vacua.
More general choices will be briefly discussed below in \S3.1.

There are two important points to emphasize about our whole discussion of
heterotic vacua: 

\noindent i) The conditions we've discussed above are sufficient to 
guarantee a solution to all orders in the string worldsheet sigma model
coupling $\alpha^\prime$.  Nonperturbatively in $\alpha^\prime$ we must
also worry about so-called ``worldsheet instanton'' effects \cite{dsww}. 
In certain classes of examples, it is known that such effects do not 
spoil conformal invariance \cite{SW}.

\noindent ii) The 
heterotic $SO(32)$ and type I theories both have 10d N=1 Super-Yang
Mills coupled to $SO(32)$ gauge fields as their low energy limits.  They
are in fact identical theories, equivalent under the strong/weak duality 
transformation \cite{typeI} 
$$
g^{het}_{MN} ~=~\lambda^{het}g^{I}_{MN},~~\lambda^{het} = {1\over \lambda^{I}}
$$
where $\lambda$ denotes the string couplings.  In particular, this implies
that after compactification on a Calabi-Yau threefold $M$,
$$
(e^{-R^2\over \alpha^\prime})_{het} \rightarrow (e^{-R^{2}\over{\alpha^\prime
\lambda_I}})_I
$$
Here, $R^2$ is the size of some rational curve in $M$.
So heterotic worldsheet instanton effects map to type I spacetime
instanton effects!  One should think of the heterotic string as the
Dirichlet 1-brane of the type I theory, so in particular the 1 brane
wrapping a holomorphic curve in a D5 brane is a ``spacetime'' instanton
effect in the D5 brane worldvolume quantum field theory \cite{douglas}. 

\subsection{M-theory}

One can get 4d N=1 theories from M-theory by compactifying on a 
7-manifold of $G_2$ holonomy.  Examples of such theories have been
studied \cite{G2} but not much is known about their physics.

\subsection{F-theory}

One gets a 4d N=1 theory by compactifying F-theory \cite{VafaF} on a Calabi-Yau
fourfold.  We have two comments to make about such theories:

\noindent i) The type IIA string has an interaction \cite{Anom} 
$$\delta S ~=~- \int B \wedge X_8(R)
$$
where B is the NS-NS 2-form and $X_8$ is a quartic polynomial in
the curvature.  So if one
compactifies on a fourfold $X$ with
$$
I = -\int_{X} X_{8}(R) \neq 0
$$
then naively there is a vacuum destabilizing tadpole for the $B$ field.
This lifts to a similar term in F-theory involving the 4-form gauge field
which couples to the SL(2,Z) invariant three-brane.  One can show that
$$
I = {\chi \over 24}
$$
for $X$ a Calabi-Yau.  In order to cancel the tadpole, one can include 
$I$ threebranes localized on $X$ but filling the transverse spacetime
$R^{3,1}$ \cite{SVW}.\footnote{Note that if $I$ is negative, there is
no way to cancel the tadpole and preserve supersymmetry.}

\noindent ii) It follows from the adiabatic argument and the duality
between the heterotic string on $T^2$ and F-theory on elliptic $K3$
surfaces that there are also dualities between 4d N=1 heterotic and
F-theory compactifications.  Suppose $X$ is an elliptic fourfold with base 
$B$, and $B$ is fibered over $B^\prime$ with $P^1$ fibers.  Then 
the adiabatic argument suggests that F-theory on $X$ is dual to
the heterotic string on the elliptic Calabi-Yau threefold $Y$ with 
base $B^\prime$.  The $I$ three-branes present in the F-theory
compactification map, in the heterotic theory, to $I$ 5-branes wrapped
on the elliptic fiber of $Y$. 

\section{More About Classical N=1 Heterotic Theories}
 
For most of these lectures, our focus will be on understanding
heterotic compactifications on Calabi-Yau threefolds.
The first step is to obtain a proper understanding of the classical
N=1 heterotic vacua.  We discussed the classical geometry briefly
in \S2.1, but in string theory the classical vacua are really
worldsheet conformal field theories.  These can predict interesting
``quantum geometrical'' phenomena when some length scales in the
compactification geometry shrink to the string length. 

The worldsheet data required to specify a ``realistic'' $SO(32)$ 
heterotic
string vacuum of the form $M_{4} \times M$ where $M_4$ is 
Minkowski space and $M$ is some compact manifold is:

$\bullet$ A free theory with left and right central charges
$c_{L} = 4, c_{R} = 6$ for the $M_4$ CFT.

$\bullet$ A $c_{L} = 16-N$ left-moving CFT composed of 
free fermions, representing the $SO(32-2N)$ current algebra.

$\bullet$ A $c_{L} = 6+N, c_{R} = 9$ conformal theory
representing the ``internal'' Calabi-Yau sigma model with
(0,2) supersymmetry.  The left moving
fermions couple to an $SU(N)$ bundle
$V \rightarrow M$. 

These are the basic structures we need.  To see if the classical
string theory differs in any interesting way from the classical
geometry, we need a considerably more precise description of the
Calabi-Yau piece of the conformal field theory.

\subsection{Gauged Linear Sigma Models}

It is prohibitively difficult in most cases to directly construct the 2d
quantum field theory as an exact conformal field theory.  What one
can do, following Witten \cite{Phases}, is construct a (0,2) 
gauged linear sigma model that should be taken under RG flow
to the desired (0,2) conformal field theory in the infrared. 
The parameters in this gauged linear sigma model
then represent the spacetime moduli of the string theory.

We will now try to construct a (0,2) supersymmetric QFT that
will
flow, in some limit, to a sigma model describing strings
propagating on a complete intersection Calabi-Yau in a
weighted projective space.
The relevant (0,2) superspace has supercharges $Q^+$ and
$\overline{Q}^+$ and SUSY multiplets:
$${\rm Chiral:}~\Phi = \phi + \theta^+ \psi + \theta^+ \bar \theta^+ 
\partial_{\bar z}\phi
$$
$$
{\rm Fermi:}~\Lambda = \lambda + \theta^+ l + \theta^+ \bar \theta^+
\partial_{\bar z} \lambda
$$
where $z, \bar z$ are complex worldsheet coordinates and the $\theta$s
are the fermionic superspace coordinates.  
In such a superspace, we
consider a $U(1)$ gauge theory the following field content. 
We take $Q+4$ chiral multiplets $\Phi^i$ with gauge charges 
$w_i$ and one additional chiral multiplet $P$ with charge $-m$.
We also take $N+1$ fermi multiplets $\Lambda^a$
with charges $n_a$ and $Q$ fermi multiplets $\Sigma_j$ with charges
$-d_j$.   
We will see momentarily that there are constraints on
the various charges, following from the conditions discussed in \S2.1. 

In addition to normal kinetic terms for all of the fields, we will include
two special terms in the action:

\noindent 1) A Fayet-Iliopoulos D-term for the U(1) gauge field
$$
S_D = r \int d^{2}z D - i{\theta \over 2\pi} \int d^{2}z f
$$
where $D$ is an auxiliary field in the U(1) gauge multiplet and
$f$ is the U(1) field strength.

\noindent 2) A (0,2) superpotential
$$
S_W = \int d^{2}z d\theta^+ \Sigma_j W^j(\Phi) + P \Lambda^a F_{a}(\Phi)
$$
where U(1) gauge invariance dictates that $W^j$ should be a polynomial
in the $\Phi$s with charge $d_j$ and $F_{a}$ should have charge
$m-n_a$.

With this Lagrangian, integrating out auxiliary fields leads to a
potential energy
$$U(\phi) = \sum_{j=1}^{Q} \vert W_{j}(\phi)\vert^{2}+ \vert p \vert^2
\sum_{a} \vert F_{a}(\phi)\vert^2 $$ 
$$+ {e^2 \over 2}( \sum_{i=1}^{Q+4}
w_i \vert \phi_i \vert^{2} - m \vert p \vert^2 - r)^{2}$$ 
The infrared behavior of the quantum field theory will be governed
by the vacua of $U(\phi)$, at least for large $\vert r\vert$ when
the excitations around the minima are very massive.
There are two such regimes:
\medskip

\noindent (I) The Calabi-Yau Phase
\medskip 

Take $r>>0$.  Assume $F_a = W_j = 0 \rightarrow \phi_i = 0$; we will
explain this momentarily. 
Then the locus of $U(\phi) = 0$ is
$$p = 0, ~~\sum_i w_i \vert \phi_i \vert^2 = r,~~W_j(\phi) = 0
$$
Dividing $$\sum_i w_i \vert \phi_i \vert^2 = r$$ 
by the $U(1)$ symmetry
leaves a copy of the $Q+3$ dimensional weighted complex projective space
with weights $w_{1},\cdots,w_{Q+4}$, and with Kahler class
(or ``size'') $r$.  Then
$W_j(\phi) = 0$ defines a complete intersection Calabi-Yau manifold
$M$ in this weighted projective space, as long as we require 
$\sum d_j = \sum w_i$.  

Further examination of the Yukawa couplings of the worldsheet 
fermions, as in \cite{DK}, reveals that the massless
combinations of the right-movers $\psi_i$
transform as sections of $TM$, while the left-movers transform
as sections of a holomorphic bundle $V\rightarrow M$, with
holomorphic structure defined by the $F_a(\phi)$.  
More precisely, this bundle is defined as the kernel of an exact
sequence
$${0\to V\to \bigoplus_{a=1}^{N+1} \CO(n_a)  {\buildrel
{\otimes F_a(\phi)}\over{\hbox to 30pt{\rightarrowfill}}}  \CO(m)\to 0}
$$
We now recognize that 
$$W_{j}(\phi) = F_{a}(\phi) = 0 \rightarrow \phi_i = 0$$
is necessary simply for the geometry of $V \rightarrow M$
to be smooth.\footnote{In fact, as long as the $F_a(\phi)$ are sufficiently
generic, mild singularities of $V$ or rather drastic singularities of $M$
may be present without adversely affecting the perturbative string theory.}
Requiring
that $V$ satisfy
the constraints of \S2.1\ then constrains the $U(1)$ gauge charges of
various worldsheet multiplets:
$$c_{1}(V) = 0 \rightarrow m = \sum n_{a} $$ 
$$c_{2}(V) = c_{2}(TM) \rightarrow $$ 
$$ m^2 - \sum n_{a}^2 = 
\sum d_j^{2} -
\sum w_i^2$$ 
In fact, both conditions are simply anomaly cancellation conditions on
the worldsheet \cite{Phases,DK}.  The first requires that there be
a non-anomalous U(1) global symmetry acting on the left movers,
while the second is simply the condition that the worldsheet U(1) gauge
anomaly should vanish! 

In this framework, the spacetime moduli fields appear as follows:

$\bullet$ Changes of $(r,\theta)$ correspond to varying the
(complexified) Kahler moduli

$\bullet$ Polynomial deformations of $W_j(\phi)$ correspond to 
moduli of the complex structure of $M$

$\bullet$ Polynomial deformations of $F_a(\phi)$ correspond to
moduli of the holomorphic bundle $V$

So for $r>>0$ we recovered the Calabi-Yau nonlinear sigma model,
with the parameters (moduli) we expect of the (0,2) CFT.  This limit
reproduces what we expect from classical geometry.  What about
the other phase, with $\vert r \vert >> 0$ but $r << 0$, which
can also be studied semiclassically?
\medskip

\noindent (II) The Landau-Ginzburg Phase

\medskip
For $r<<0$, the minimum of $U(\phi)$ occurs when 
$\vert p\vert^2 = -{r\over m}$
and $\phi_i = 0$.  The U(1) gauge symmetry is broken to
a $Z_m$ by $\langle p \rangle$, which carries nonzero U(1)
charge.  The massless degrees of freedom are the
$\phi$s, $\Sigma$s, and $\Lambda$s, governed by
a superpotential
$$W = \Sigma^j W_j(\phi) + \Lambda^a F_a(\phi)
$$
The unbroken $Z_m$ discrete gauge symmetry instructs us
to take a $Z_m$ orbifold of the naive Landau-Ginzburg theory,
to obtain a so-called ``Landau-Ginzburg orbifold.'' 

\vskip 1cm
$$
\vbox{
{\centerline{\epsfxsize=3in \epsfbox{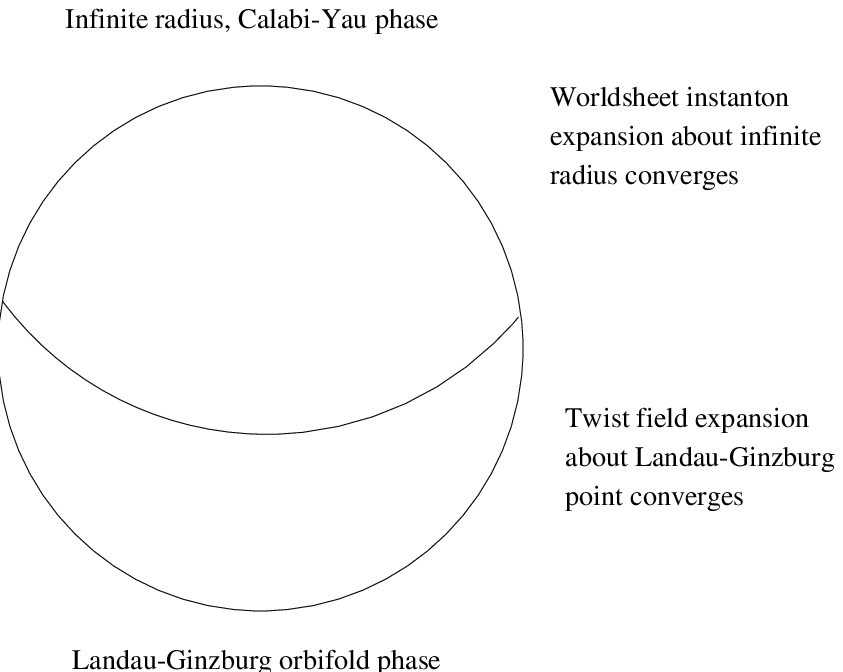}}}
{\centerline{ FIGURE 1:
A picture of the Kahler moduli space.}}
}
$$
\vskip .5cm

So the global picture of the Kahler moduli space, controlled
by $(r,\theta)$, is roughly as follows.
As $r$ decreases from $\infty$ to $0$, the Calabi-Yau sigma
model coupling gets stronger and stronger until finally the
perturbation series for various correlators diverges.  
Going ``past'' this point, to negative $r$, one finds that
there is a $\it new$ convergent perturbation expansion, 
this time in twist fields about the Landau-Ginzburg orbifold,
which becomes more and more weakly coupled as $r\rightarrow
-\infty$.  This is summarized in Figure 1 above.

\subsection{A Glimpse of Quantum Geometry}

The Landau-Ginzburg phases are as ``far away'' from the 
$r \rightarrow \infty$ classical geometry limit as we can get
in the Calabi-Yau moduli space.  One might therefore expect
that examples of ``stringy'' quantum geometrical effects might
be particularly manifest there.  One such phenomenon, which
occurs in string theory but not classical geometry and is 
mediated through the Landau-Ginzburg phase, is a certain
kind of smooth topology change \cite{DK,DKtwo,CDG}.

Notice at the Landau-Ginzburg point, there is no invariant
distinction between the $\Sigma_j$s and the $\Lambda^{a}$s (or
the $W_j$s and the $F_a$s).  
This is in marked distinction to the distinguished roles these
fields play at large radius, where the $W_j$s determine $M$ and
the $F_a$s determine the holomorphic structure of $V$. 
This makes it possible for us to imagine an automorphism which
exchanges certain $\Sigma$s and $\Lambda$s, while at the same
time exchanging certain $W$s and $F$s, at the Landau-Ginzburg
point.  
Since the spectrum of chiral fields remains invariant under such
a re-labeling, any models related by such a symmetry must 
(at large radius) be
Calabi-Yau complete intersections in the $\it same$ weighted
projective space.
Examples of such automorphisms yield pairs of
topologically distinct Calabi-Yau manifolds $M, X$ 
(which can even have different Euler characters $\chi(M) \neq
\chi(X)$) 
and vector bundles 
$V \rightarrow M$ and $E \rightarrow X$ 
which are smoothly connected through their
(shared) Landau-Ginzburg phase.\footnote{In such examples, 
$\int_M c_{3}(V) = \int_{X} c_{3}(E)$; the net spacetime
chirality does not change.}  
The Landau-Ginzburg phase then
plays the role of a multi-critical point in the CFT moduli space. 
One can turn on distinct moduli, which are not mutually
integrable,
to go to large-radius in either of the two distinct Calabi-Yau
phases. 
Simple concrete examples of this phenomenon are
discussed in \cite{DK,DKtwo,CDG} and we will give no further
details here.

\vskip 1cm
$$
\vbox{
{\centerline{\epsfxsize=3in \epsfbox{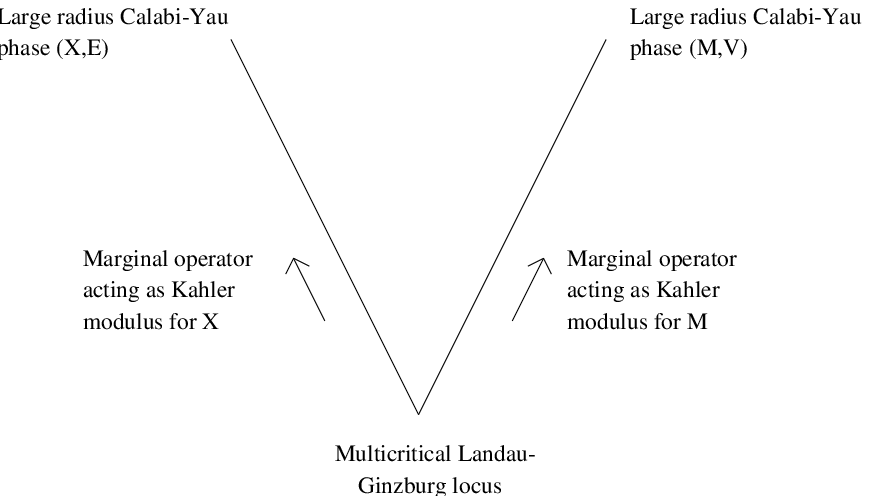}}}
{\centerline{ FIGURE 2:
Topology change through LG phase.}}
}
$$
\vskip .5cm

\section{General Remarks about Singularities} 

As discussed by many other lecturers, SUSY field theories and
string theories typically have moduli spaces ${\cal M}$ 
of physically
inequivalent vacua parametrized by scalar VEVs $\langle \phi
\rangle$.  These moduli spaces ${\cal M}$ typically have special
points $m^* \in {\cal M}$ where the physics is $\it singular$. 
Interesting dynamics occurs there.  A lesson of the recent
advances is that understanding what happens at singular points
in the passage from the classical to the quantum theory is
the crucial step in ``solving'' the quantum theory.

\vskip 1cm
$$
\vbox{
{\centerline{\epsfxsize=3in \epsfbox{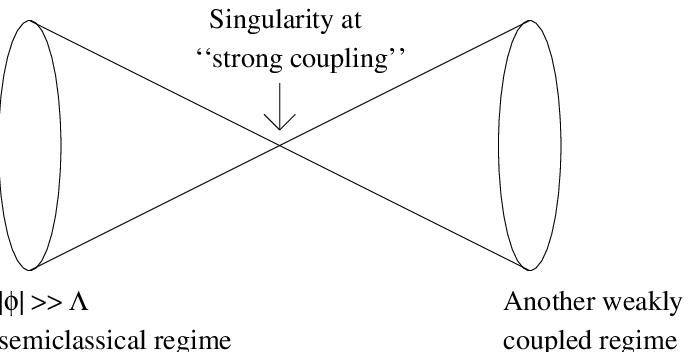}}}
{\centerline{ FIGURE 3:
A depiction of some ``typical'' moduli space}}
{\centerline {in a theory with scale $\Lambda$.}}
}
$$
\vskip .5cm

\subsection{Some Examples}          
 
Let us begin by reviewing some prototypical examples.  At least four
sorts of phenomena have been observed in 4d gauge theory and string
theory: 

\noindent 1) A classical singularity is $\it removed$ in the 
quantum theory.

\noindent Example: Consider the $SU(2)$ gauge theory with
$N_F = 2$ flavours of quarks, i.e. four doublets $d_i$,
$1 \leq i \leq 4$.  The coordinates on the classical moduli space
${\cal M}_{cl}$ are $V_{ij} = \epsilon_{\alpha \beta}d_i^{\alpha}
d_j^{\beta}$.  The equation defining ${\cal M}_{cl}$ is
$$
{\cal M}_{cl}:~Pf(V) = \epsilon^{ijkl}V_{ij}V_{kl} = 0
$$
In the quantum theory, this is deformed to
$$
{\cal M}_{qu}:~Pf(V) = \Lambda^{4}
$$
where $\Lambda$ is the scale of the gauge theory.  The singular
point $V_{ij} = 0$ on the classical moduli space, where the $SU(2)$
gauge symmetry is restored, is $\it removed$ in the quantum theory
\cite{Seiberg}. 

\noindent 2) The mathematical structure of ${\cal M}_{cl}$ and
${\cal M}_{qu}$ agrees, but the physical interpretation of the
singular points changes.

\noindent Example: Consider $SU(N_c)$ gauge theory with $N_F = 
N_C + 1$ flavours of quarks.  Then the classical singular points
are due to the appearance of massless gluons (``unHiggsing''), while
the quantum singularities are due to the appearance of additional
massless composite ``baryons'' and ``mesons'' \cite{Seiberg}.

\noindent 3) There are still singularities on ${\cal M}_{qu}$,
but the mathematical structure (and hence also the physical
interpretation) is modified.

\noindent Example: Take the famous case of N=2 supersymmetric pure
$SU(2)$ gauge theory.  The N=2 gauge multiplet contains a complex
scalar $\phi$ in the adjoint of $SU(2)$.  There is a one complex
dimensional moduli space of vacua, parametrized by the gauge invariant
coordinate $u = Tr (\phi^{2})$.  In the classical moduli space, there
is a singular point at $u = 0$ where the $SU(2)$ gauge symmetry
is restored.  In the quantum theory, this splits into two singular
points at $u = \pm \Lambda^2$ (where $\Lambda$ is the scale of
the theory), where a monopole and a dyon become massless.  This is
the crucial insight needed to solve the theory \cite{SeiWit}.  

\vskip 1cm
$$
\vbox{
{\centerline{\epsfxsize=3in \epsfbox{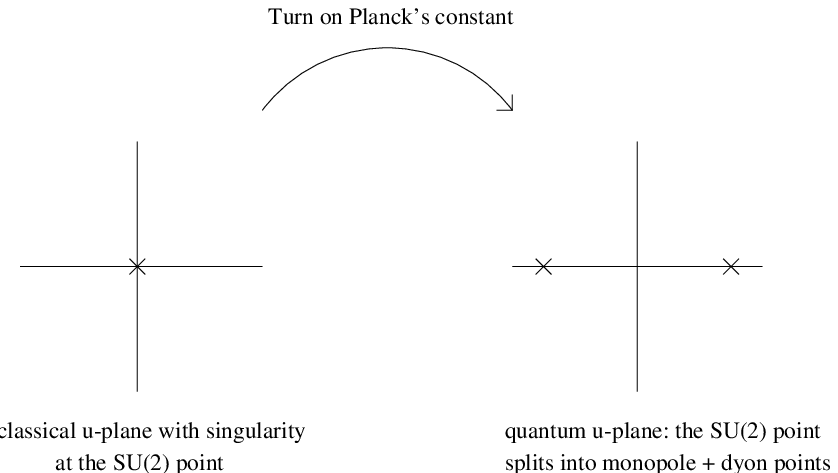}}}
{\centerline{ FIGURE 4:
Pictures of the u-plane}}
{\centerline{in the classical and  
quantum theory.}}
}
$$
\vskip .5cm

There is a $\it fourth$ possibility which seems to occur in string
theory.  Namely:

\noindent 4)  The singularity is `inexplicable' classically.  In the
quantum theory, it arises due to quantum loops or nonperturbative
dynamics of solitons.  These (quantum) effects are independent of
$g_{string}$ because of the unusual couplings of these solitons.

\noindent Example:  A famous example is provided by Strominger's
resolution of conifold singularities in IIB compactifications on
Calabi-Yau threefolds \cite{andy}.  There, $M$ is a 
Calabi-Yau and $X \subset M$ a three-cycle.  There is a scalar
component $z$ in a Ramond-Ramond $U(1)$ vector multiplet given by
$$
z = \int_{X} \Omega
$$
One can wrap a Dirichlet three-brane on $X$ to obtain a BPS state
whose mass is bounded by $\vert \int_{X} \Omega \vert$, which is
just the size of the 3-cycle $X$.  In particular, as $z \rightarrow
0$ one obtains a new massless charged particle, arising from the wrapped
3-brane!   
This explains a log singularity in the $\it classical$ IIB
prepotential.
We will find even more dramatic examples of such phenomena in
N=1 compactifications, where $\it nonperturbative$ dynamics of
solitons will explain singularities of classical string theory.

We should note that a physical interpretation of the singular points
is necessary for consistency of the theory.  But more importantly, it
allows one to gain control over novel physical effects.  For instance:

$\bullet$ In 4d N=2 models, the quantum understanding of conifold
points gives a physical realization to the 
idea of ``conifold transitions.''  These allow one
to connect up the moduli spaces of all Calabi-Yau compactifications
of type II strings \cite{old,gms}! 

$\bullet$ In the 4d N=1 case, we will see later on that 
singular points allow one to connect up models with different
net numbers of chiral generations \cite{chiral}!

There is yet another motivation for studying singularities, which
is generic to N=1 supergravity models with superpotentials.

\subsection{Poles and Superpotentials}

It turns out that in 4d N=1 models, singular points are
$\it necessary$ in order to have a non-vanishing superpotential
\cite{bw}.

Recall that N=1 supergravity theories are specified by a Kahler
potential $K(\Phi, \Phi^*)$ as well as a holomorphic superpotential
$W(\Phi)$ and gauge coupling function $f(\Phi)$.  

$\bullet$ The Lagrangian ${\cal L}$ has a Kahler invariance under
which 
$$
K(\Phi, \Phi^*) \rightarrow K + F(\Phi) + F^*(\Phi^*)
$$
This Kahler transformation must be accompanied by a chiral
transformation on the fermion fields in $\Phi$ and the gravitino. 

$\bullet$ In the presence of a superpotential $W$, the Kahler
invariance still holds if in addition to the shift of $K$, one has
$$
W \rightarrow e^{-F}W
$$
This means that W is actually a section of a line bundle over the
moduli space ${\cal M}$ parametrized by the $\Phi$s.

$\bullet$ The $\it connection$ on this bundle has covariant
derivatives
$${DU \over {D\phi}} = {\partial U \over {\partial \phi}} + 
{\partial K\over {\partial \phi}} U$$ 
$${DU \over {D\phi^*}} = {\partial U \over {\partial \phi^*}}$$ 
from which it follows that its $\it curvature$ is
$- {\partial^{2}K\over{\partial \phi \partial \phi^*}} = - g_{\phi \phi^*}$.
Since to get the right sign of the kinetic energy we want the metric
on ${\cal M}$ to be positive definite, this means that the superpotential
is a section of a line bundle of $\it negative$ curvature over ${\cal M}$. 

But now a standard theorem of complex geometry tells us that
if ${\cal M}$ is compact, or can be suitably compactified, then
as a section of a line bundle of negative curvature $W$ $\it {must~ have}$
poles if it does not identically vanish \cite{bw}.
This means that

\noindent a) The known tree level superpotentials in string theory
must have poles at codimension one.  We must explain these poles
physically.

\noindent b) If we want to find ``non-perturbative'' superpotentials
which e.g. dynamically break supersymmetry or destabilize would-be
supersymmetric ground states, they too must have poles.  So it is
sometimes easiest to see if such a superpotential is generated by
going to the potential singular points in moduli space (which are often
clearly identifiable in a given model) 
and checking there. 

\section{The Simplest Singularities of Heterotic Strings on K3}

Our eventual goal is to study singularities of 4d N=1 compactifications
of the heterotic string.  However, our strategy will be to use knowledge
of singular behavior in $\it six$ dimensions, fibered over an additional
sphere, to determine the four-dimensional physics.  So we first need
to review the phenomena which accompany singularities in 
6d (0,1) supersymmetric heterotic string compactifications.
We will study the simplest cases here; more complicated singularities
in (0,1) vacua have recently been discussed in e.g. \cite{newsix}. 
 
\subsection{Heterotic SO(32) on K3}

When we compactify the $SO(32)$ heterotic string on a K3 to
obtain a 6d $N=1$ theory, the Bianchi identity
$$
dH = Tr(R\wedge R) - Tr(F \wedge F)
$$
together with 
the fact that $\int_{K3} Tr (R\wedge R) = 24$ tells us that
we must put 24 Yang-Mills instantons in the $SO(32)$.  
The natural place to look for singularities is then when one
of the instanton scale sizes shrinks to zero. 
The novel physics of such points was explained by Witten 
\cite{small}.

What can $\it possibly$ happen at such a point?          
Singular points in the moduli space are most simply explained as
points where new particles are becoming light.  In 6d theories
with (0,1) supersymmetry, the only massive supermultiplet with
spins $\leq 1$ is the massive vector multiplet.  As its mass
goes to zero, it becomes a massless vector + hypermultiplet.
So the only relevant $\it free$ low energy dynamics is that
the Higgs mechanism might be turned on or off.   

Consider a shrinking instanton in $SO(N)$ then.  Since it is 
small the details of the $K3$ are irrelevant, so we might
as well pretend it is on $R^4$.  Then 
$$dim({\cal M}_{inst}) = 4N - 8 $$ 
$${\cal M}_{inst} = {\cal M}^\prime_{inst} \times R^4 $$ 
$$dim({\cal M^\prime}_{inst}) = 4N - 12 $$ 
where the $R^4$ factor is just the instanton center of mass.
Now if one proposes to get ${\cal M}^\prime_{inst}$ by
Higgsing a gauge group $G$ with $k$ hypermultiplets, then
$$
dim ({\cal M}^\prime_{inst}) = 4 (k-dim (G))
$$
So one needs $k - dim(G) = N-3$.  Hence, it is natural
to conjecture that $G=SU(2)$ and the matter is N 
half-hypermultiplets in the $\bf{2}$ of $SU(2)$.  

At least two strong pieces of evidence favor the conjecture
that a nonperturbative $SU(2)$ together with an $\bf{(N,2)}$
of $SO(N) \times SU(2)$ half-hypermultiplets appear when an
$SO(N)$ instanton shrinks to zero size.

\noindent 1) There is a well known ADHM construction of
instanton moduli space \cite{ADHM}.  The moduli space of the proposed
nonperturbative gauge theory precisely reproduces this
construction.

\noindent 2) There is a description of instantons at scale
sizes $\rho >> l_{s}$ on $K3$ as the (0,4) superconformal
field theory of ``solitonic fivebranes'' of thickness $\rho$.
As $\rho \rightarrow 0$, the fivebrane goes to zero thickness.
Since this whole process is independent of $g_{string}$, we
can take $g_{string} \rightarrow \infty$.  Then by heterotic/
type I duality one obtains a weakly coupled type I string.
It is natural to think that the zero thickness object you
have sitting at a point on $K3$ is now nothing but a 
type I Dirichlet 5-brane!  The properties of the D5 brane
are in complete agreement with the previous conjecture.

One can similarly see, as is clear from 2) above, that
shrinking $k$ instantons at the same (smooth) point yields
an $Sp(k)$ gauge theory.

\subsection{Singularities of Heterotic $E_8 \times E_8$ on
K3}  

The $E_8 \times E_8$ heterotic string on $K3$ can be thought of
as M theory on ${S^1/Z_2} \times K3$ \cite{hw}.
Naively, the Bianchi identity for the $H$ field tells one that
one must place $n_{1,2}$ instantons in the two $E_8$s at the
ends of the world, where
$$n_{1} + n_{2} = 24
$$
If for simplicity we put the $n_{1,2}$ instantons in an $SU(2)$
subgroup of each $E_8$, then the unbroken gauge symmetry on
each end of the world is $E_7$, and the charged matter consists of 
$\vert {1\over 2}n_{1,2} - 2 \vert$ $\bf {56}$s. 
One can now ask, what happens as one of these $E_8$ instantons
shrinks to zero size?

In \S5.1, we saw that the only possible $\it free$ IR dynamics
consistent with (0,1) supersymmetry is that the Higgs mechanism
can be turned on or off. 
This insight sufficed in determining the behavior of the $SO(32)$
small instanton.
However, it is easy to convince oneself
that the singularity in the $E_8$ case must involve more subtle
behavior than this (a quick glance at the dimensions of $E_8$
representations makes this clear).  In fact, as explained in
\cite{GanHan,SWstring}, the six-dimensional theory at the end of
the world is governed by a nontrivial fixed point as an 
instanton shrinks
to zero size.  This is a multicritical point.  The zero-size instanton
constitutes a fivebrane at the end of the world.  There are two
phases connected by the fixed point theory: the Higgs phase is
obtained by enlarging the small instanton, while a distinct
Coulomb-like phase is obtained by moving the fivebrane off into
the eleven-dimensional bulk.  The Bianchi identity is generalized
to \cite{dmw} 
$$n_{1} + n_{2} + n_{5} = 24
$$
where $n_5$ is the number of fivebranes.  It follows from the
previous discussion that the number of 1/2 ${\bf 56}$s at
the end of the world the fivebrane leaves decreases by one.

The fivebrane worldvolume theory contains a tensor multiplet
consisting of a real scalar, an antisymmetric tensor with selfdual
field strength, and fermions.  The real scalar's VEV gives the
distance of the fivebrane from the end of the world.  The
perturbative heterotic theory has one tensor multiplet, whose
two-form potential combines with that in the 6d gravity multiplet
(which has anti self-dual field strength) to form an unconstrained
antisymmetric tensor field.  The transition back to the perturbative
heterotic theory from the phase with extra tensor multiplets cannot
be described in conventional Lagrangian field theory, since the
extra tensors cannot pick up masses without having any anti self-dual
partners to couple to.

\vskip 1cm
$$
\vbox{
{\centerline{\epsfxsize=3in \epsfbox{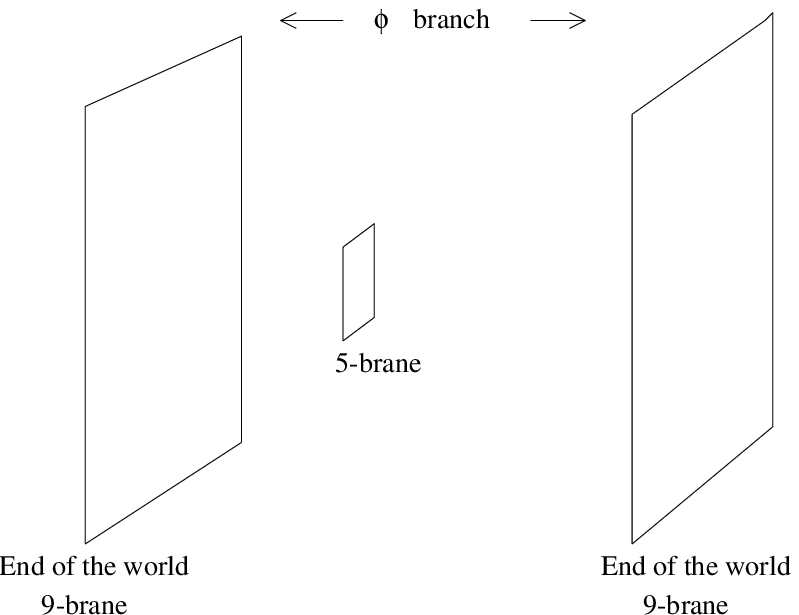}}}
{\centerline{ FIGURE 5:
A fivebrane in bulk.}}
}
$$
\vskip .5cm

If we consider a single fivebrane and call the real scalar in
the tensor multiplet $\phi$, then there are noncritical strings
of tension $\phi$ and ${1\over \alpha^\prime} - \phi$ arising
from membranes 
stretched between the fivebrane and the two ends of the world.
As the fivebrane approaches an end of the world, the tension of
the associated noncritical string goes to zero.  For this reason, the 
nontrivial fixed points which occur when an $E_8$ instanton
shrinks to zero size are sometimes called ``tensionless
string theories.''
These fixed points (and their close relatives, obtained by compactifying
on circles or tori) have been extensively investigated, in e.g.
\cite{seibetal}.

\section{K3 Fibrations and $SO(32)$ Strings}

So far, we have studied the simplest singularities of the heterotic
theories compactified on K3 to six dimensions.  Our real interest is
to use this knowledge to understand singularities in the 4d N=1
theories coming from compactifications on Calabi-Yau threefolds which
are K3 fibrations.  If we study singularities which occur 
at a point in the
generic K3 fiber, then we should be able to use our knowledge of
the six-dimensional physics to learn about the four-dimensional
physics.  We proceed to do this for the $SO(32)$ string in this
section, and discuss the $E_8 \times E_8$ string in \S7. 

\vskip 1cm
$$
\vbox{
{\centerline{\epsfxsize=3in \epsfbox{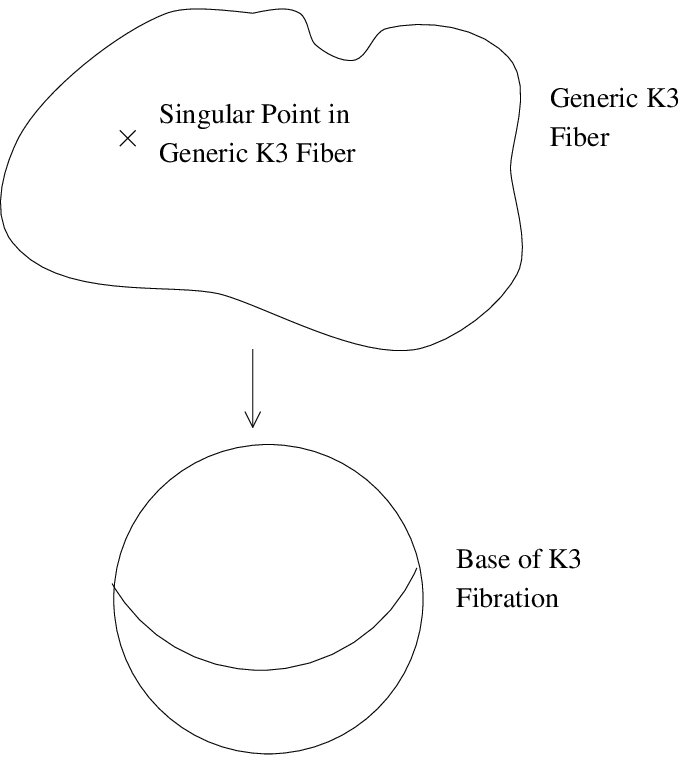}}}
{\centerline{ FIGURE 6:
A K3 fibration with}}
{\centerline{singularity in the generic  
fiber.}}
}
$$
\vskip .5cm

The simplest situation which we are equipped to study is the
case that the K3 fibration $M$ has a single small instanton in 
the generic fiber.  Then the generic fiber theory has an enhanced
nonperturbative $SU(2)$ symmetry as described in \S5.1.  We will
now see how this ``fibers'' to yield the four-dimensional story.

\subsection{Finding the 4d Gauge Theory}

The question we need to answer is: What of the $SU(2)$ gauge theory
with 32 1/2 $\bf{2}$s of hypermultiplets survives in the
4d N=1 theory?  The method we will use is to analyze 
adiabatically by making the $P^1$ base $C$ large and 
compactifying the 6d theory on $C$ with a twist, to preserve
precisely 4d N=1 supersymmetry \cite{kss}. 

The Lorentz group in the ``K3 part'' of the theory decomposes
as $SO(4) \rightarrow SU(2)_H \times SU(2)^\prime$, where
$SU(2)_H$ is the holonomy group of K3.  The six dimensional
supercharges are in a doublet of $SU(2)^\prime$, with
$J_{3}^\prime = \pm 1/2$.  
The Lorentz group of $C$ is a $U(1)$, and
each 6d supercharge decomposes into pieces with $q=\pm 1/2$ under
this $U(1)$.  So if we twist and call the Lorentz $U(1)$
$q + J_{3}^\prime$, we will end up keeping precisely half
of the six dimensional supersymmetry.

The four-dimensional $SU(2)$ gauge bosons also survive the reduction
as constants on $C$.  So we will still get a nonperturbative
$SU(2)$ gauge theory.  The real question is, how many doublets
survive?

For the gauge bundle $V \rightarrow M$ an $SU(N)$ bundle, 
the $\bf{(32,2)}$ splits into 2N fields which couple to
$V$ or $V^*$ and 32-2N which couple to the trivial bundle.
To count how many $\it fermions$ survive (the bosonic spectrum
follows by SUSY), we just need to count the number of zero modes
of the Dirac operator on $C$.  This counting is possible using
several simple observations: 

\noindent $\bullet$  
Zero modes of $\dsl + iA$ on $C$, where $\dsl$ is the
covariant Dirac operator and $A$ is the gauge connection
on $V$, are in one to one correspondence with  
elements of $H^{0}({\cal O}(-1) \otimes V)$.  Here
${\cal O}(n)$ simply denotes the line bundle on $C$ with
first Chern class $n$, and ${\cal O}(-1)$ is the spin bundle.

\noindent $\bullet$ It is also true that $\it any$ 
holomorphic bundle $V$ splits as a direct sum of line bundles
when it is restricted to $C$
$$V\vert_{C} = \sum_{i=1}^{N} {\cal O}(a_i)
$$

\noindent$\bullet$ Finally, to count the surviving fermions
one needs to know that $h^{0}({\cal O}(n)) = n+1$ for 
$n \geq 0$ and vanishes for $n<0$.

Putting these facts together, we find that the 32-2N $\bf{2}$s which
couple to the trivial bundle do $\it NOT$ survive, since
$H^{0}({\cal O}(-1))$ is trivial.  On the other hand, the
2N doublets which couple to $V$ and $V^*$ contribute
$${\rm \# ~4d~ doublets} = h^{0}(V\otimes {\cal O}(-1)) $$ 
$$  + h^{0}(V^* \otimes {\cal O}(-1))$$ 
So on $\it curves$ of singularities of the sort I have been
describing, there is a nonperturbative enhancement of the spectrum
consisting of $SU(2)$ gauge fields with a calculable number of
doublets.  The dynamics in this extra sector should allow us to
explain various singularities of $SO(32)$ string compactifications.

Note that the coupling of the $SU(2)$ is not the normal string
coupling of the heterotic string.  Instead, it is 
$${1\over g^{2}} = {R^{2}\over \alpha^\prime} \rightarrow $$ 
$$\Lambda_{SU(2)} \sim M_{S} e^{-{R^{2}\over \alpha^\prime}}$$ 
where $R$ is the radius of the base curve $C$.
On the other hand, in the type I theory these gauge fields
are governed by the string coupling too.

\subsection{Uses of the 4d Spectrum}

There are two phenomena which occur in heterotic (0,2) models
which can be explained by these nonperturbative $SU(2)$ gauge theories.

\noindent 1) Worldsheet instanton destabilization

There can exist curves $C$ with worldsheet instantons (nontrivial
maps of Euclidean string worldsheets which wrap $C$) contributing 
$$
W \sim e^{-{\int_C J \over \alpha^\prime}} \sim e^{-
{R^2\over \alpha^\prime}}
$$
to the $\it spacetime$ superpotential \cite{dsww}.
It is of interest to know for which models such a superpotential
appears, and to give an ``explanation'' of the superpotential in
terms of conventional low-energy physics.

\noindent 2) Poles in Yukawa couplings

If we compactify the $SO(32)$ heterotic string on a K3 fibration $M$
and take $V \rightarrow M$ to be a holomorphic deformation of $TM$,
then the low energy effective theory has $SO(26) \times U(1)$
gauge group.  The charged matter content consists of
$$h^{1,1}(M) \times 26_{1} \oplus 1_{-2}$$ 
$$h^{2,1}(M) \times 26_{-1} \oplus 1_{2}$$ 
Quantities of interest in ${\cal L}_{eff}$ include the Yukawa
couplings
$$y_{ijk}26^{i}_{1}26^{j}_{1}1^{k}_{-2}$$
where $y_{ijk}$ is a function of the moduli.

It has been observed that in many models \cite{CDGP,SW}\ the
following occurs:  If $z$ is a modulus of the theory such that
at $z \rightarrow 0$ $V \rightarrow M$ becomes singular,
then $y_{ijk}$ behaves like
$$y_{ijk} \sim {1\over z} \tilde y_{ijk}$$
as $z \rightarrow 0$, where $\tilde y_{ijk}$ is finite in
the limit.  It is necessary for consistency of the theory to
find a dynamical explanation for such poles.

We will try to explain phenomena 1) and 2) in turn, for this class of
singularities. 

{\noindent {\bf Application I: Vacuum Cleaning}} 

Suppose that at codimension one in the moduli space of vacua, one finds
a singularity of the sort we've been discussing with
$$V\vert_{C} = {\cal O}(1) \oplus {\cal O}(-1) \oplus {\cal O}(0) \oplus
\cdots$$
where $\cdots$ denotes further trivial factors.  Then one gets a
nonperturbative enhancement of the spectrum consisting of $SU(2)$ gauge
theory with 2 massless doublets $d_{1,2}$.  

\noindent $\bullet$ Classically, this theory has a 1 complex
dimensional moduli space of vacua with $V_{12} = \epsilon_{\alpha
\beta}d_{1}^{\alpha}d_{2}^{\beta}$ being the (gauge invariant)
coordinate.  

\noindent $\bullet$ Quantum mechanically, Affleck, Dine and Seiberg
showed some years ago \cite{ads} that a one instanton effect in this theory
dynamically generates a superpotential
$$W_{dyn} = {\Lambda^{5}\over V_{12}}$$
where for us $\Lambda \sim M_{S}e^{-{R^{2}\over \alpha^\prime}}$.
That is exactly the sort of superpotential
we expect worldsheet instantons to generate in heterotic string
theory!  There are several checks:

i) The pole at $V_{12}=0$ 
can play the role of the pole required (as reviewed in 
\S4.2) for a nonzero
superpotential in supergravity.

ii) The fact that $C$ is a curve of $\it singularities$ agrees
perfectly with independent analysis of when a worldsheet instanton
superpotential can be generated \cite{SW}.

iii) By zero-mode counting arguments in an explicit worldsheet
instanton computation, one can show \cite{Dist}\ that this splitting
type is one of only $\it two$ cases where a 
worldsheet instanton superpotential is possible.
This involves a counting of heterotic string worldsheet fermion zero
modes instead of nonperturbative $SU(2)$ doublets! 

The other splitting type consistent with a worldsheet instanton
generated superpotential is trivial splitting.  From our analysis,
this case corresponds to a pure $SU(2)$ gauge theory, and gaugino
condensation occurs, which again would generate a potential
$W \sim \Lambda^3$ in the gauge theory analysis.  In this case,
since there are no moduli which remove the $SU(2)$, the compactification
would be generically singular, however.

So, we have found that there is a complete agreement between
situations where the nonperturbative $SU(2)$ gauge theory can
lead to vacuum destabilization and situations where worldsheet instanton
considerations allow a superpotential to be generated \cite{ks}.
Three comments are in order:

\noindent 1) Duality maps these worldsheet instanton effects in the
heterotic string to both spacetime instantons of the type I theory
(in the worldvolume gauge theory on a 5-brane), and Euclidean 5-branes
wrapping divisors of arithmetic genus 1 in F-theory on a dual
Calabi-Yau fourfold \cite{witsup}.

\noindent 2) The physical upshot is that many heterotic or type I
vacua with no nontrivial dynamics in the obvious (perturbative)
gauge groups can still be destabilized by dynamics in less obvious
nonperturbative gauge groups. 

\noindent 3) The techniques of \cite{witsup}\ have been used to
study considerably more elaborate examples of nonperturbative
superpotentials, which exhibit modular behavior \cite{dgw,cl}. 

\noindent{\bf{Application II: Poles in Yukawa Couplings}}

Analyzing the situations where the generic $K3$ fiber is
singular and there is a pole in a Yukawa coupling, we find that
$$V\vert_{C} = {\cal O}(2) \oplus {\cal O}(0) \oplus {\cal O}(-2)$$
This implies by our previous analysis that at the singular point
in moduli space one finds an $SU(2)$ gauge theory with
$N_F = 2$, i.e. 4 doublets.

So how does this explain the pole in the Yukawa couplings of
$SO(26) \times U(1)$ charged matter \cite{kss}?
Denote the doublets $d_{i}, 1\leq i \leq 4$.  As we reviewed in
\S4.1, SUSY QCD with $N_F = N_C = 2$ has a smooth quantum moduli
space ${\cal M}_{qu}$ which is a deformation of the classical
one:
$${\cal M}_{qu}: Pf(V) = \Lambda^{4}$$
where $\Lambda$ is the dynamical scale of the gauge theory.
The constraint restricting the fields to lie on ${\cal M}_{qu}$
can be imposed with a Lagrange multiplier in the
superpotential
$$W_{0} = \lambda(Pf(V) - \Lambda^{4})$$

In the gauge theory, the Higgs branch has 5 complex dimensions.
In our problem, there is only $\it one$ relevant dimension in
the moduli space, parametrized by $z$.  This implies the existence
of a tree-level superpotential
$$\delta W = V_{13}^{2} + V_{14}^2 + V_{23}^2 + V_{24}^2$$
Integrating out the $V$s appearing in $\delta W$, we find
that they vanish.  Then the Pfaffian constraint obtained by
integrating out $\lambda$ reveals
$$V_{34} = {\Lambda^{4} \over V_{12}}$$
This yields the required complex 1-dimensional moduli space.

What about the pole?  Consider adding a term
$$\delta \tilde W = (26_{1} 26_{1} 1_{-2})V_{34}$$
to the superpotential.  Then altogether we have
$$W = W_0 + \delta W + \delta \tilde W$$
Now integrating out $V_{13}, V_{14}, V_{23}, V_{24}$ and $\lambda$,
we find 
$$W_{dyn} = {\Lambda^{4}\over V_{12}} (26_{1} 26_{1} 1_{-2})$$
This is the desired pole, with $z \sim V_{12}$! 

Note that nonperturbative $\it quantum$ phenomena in the $SU(2)$ gauge
theory can explain $\it classical$ heterotic string singularities
because $e^{-{R^2\over \alpha^\prime}}$ is playing the role of
$e^{-{1\over{hg^2}}}$.  This is the analogue of the special properties
of RR fields used by Strominger in explaining classical singularities
of N=2 string vacua with quantum loops of light solitons \cite{andy}.  

\section{K3 Fibrations and $E_8 \times E_8$ Strings}

We will now briefly discuss a similar analysis of singular $E_8 \times
E_8$ string compactifications on $K3$ fibrations.  This will reveal a
novel phenomenon which we did not encounter in the $SO(32)$ case.
We will basically motivate and summarize the results of \cite{chiral},
where all of the details can be found. 

One of the longest standing mysteries of string theory has been the
vacuum degeneracy problem.  It is hard to see how a dynamical selection
principle could choose between $\it disconnected$ components of
the space of vacua.  For the case of compactifications with N=2
supersymmetry, the situation improved considerably with the work of
\cite{gms}. 

The case of N=1 vacua is qualitatively different from the N=2 case.
N=1 is the first case where chiral gauge representations occur.  
No transition which occurs in weakly coupled
Lagrangian field theory can change the net number of chiral 
generations.  
Nonetheless, we will find in this section 
that in fact some vacua of the $E_8 \times
E_8$ string
with different generation numbers can be connected by phase
transitions \cite{chiral}.
These transitions involve going through a point in moduli
space where there is no $\it free$ long distance description
of the physics.

\subsection{Fibration Picture of the Phase Transition}

As in \S6, we consider compactification on a K3 fibration $M$ with
base sphere $C$.  Call the radius of $C$ $R$.  Above the scale 
$1/R$, the physics can be analyzed in terms of an effective
six-dimensional compactification on $K3$.

Suppose an $E_8$ small instanton develops in the generic fiber.
For simplicity, suppose it develops in an $E_8$ factor which is on
a locus with unbroken $E_7$.  Then as we discussed in \S5.2, moving
the small instanton fivebrane off into the bulk (i.e., turning on
the scalar in its tensor multiplet) yields a vacuum with one less
1/2 $\bf{56}$ localized on the wall it left.  
At the intersection of the two branches, the physics is governed
by a nontrivial RG fixed point. 

Let us now consider the theory below the scale $1/R$, i.e. in four
dimensions.  For definiteness let us consider an $SU(3)$ bundle
$V$ embedded in one $E_8$ on the Calabi-Yau threefold $M$.
This leads to an $E_6$ gauge symmetry in spacetime.  In a class of
examples, an analysis of the 4d spectrum reveals the 
following \cite{chiral}:
Each 6d 1/2 ${\bf{56}}$ hypermultiplet descends to a set of
${\bf 27}$s $\it or$ to a $\bf{\overline{27}}$ in the 4d theory.
There is also some matter which does not descend from matter in the
6d theory, i.e. which is not associated with the generic fiber.
This matter is therefore associated with singular fibers.

In the 4d theory, a zero-size instanton in the generic fiber corresponds
to a fivebrane wrapped around the base $P^1$.  As shown in \cite{witwrap}
it remains consistent with supersymmetry for the fivebrane to move away
from the end of the world, as long as it is wrapped on a holomorphic
curve.  In this phase, since of the the 1/2 $\bf{56}$s has been 
removed in six dimensions, the corresponding $\bf{27}$s or $\bf{\overline
{27}}$ have been removed in the four-dimensional theory.  One can argue
that no additional states from singular fibers are produced in the
transition.  So we find that the net generation number has shifted. 
This is possible because at the point of transition, there is no weakly
coupled description of the infrared physics:  The theory is governed
by a nontrivial RG fixed point obtained by ``fibering'' the 6d fixed
point over a sphere.  It would be very interesting to obtain a better 
understanding of such exotic 4d N=1 fixed points; examples have been
constructed on brane probes in \cite{aks}.

\subsection{Instanton Effects}

So far, we have discussed the 4d theory below the scale $1/R$ set
by the Kaluza-Klein modes on the sphere.  There are also
lower dynamical scales in the problem, set by the strength of worldsheet
instanton effects.  As we mentioned in \S5.2, in addition to the
fundamental string there are two noncritical strings of interest when
the fivebrane moves off into the interval.  If $\phi$ parametrizes its
position on $S^1/Z_2$, then the tension of these noncritical strings
goes like $\phi$ and ${1\over \alpha^\prime} - \phi$.  So euclidean
worldsheets of these strings wrapping $C$ will generate effects that
go like $e^{-R^2 \phi}$ and $e^{-R^2({1\over \alpha^\prime}-\phi)}$.  

The physics of these instanton effects can be classified according to
the splitting type of $V$ over $C$ at the small instanton singularity,
as in \S6.  Actually, in general there is another vector bundle $V_2$
embedded in the second $E_8$.  At the point where the fivebrane is
``absorbed'' into the second $E_8$, $V_2$ is also singular over $C$
and its splitting type is of interest.  There are a few different cases
one can consider, and we mention two of the simplest below.
Note that T-duality relates this theory, compactified on an additional
circle, to the $SO(32)$ theories we discussed in the previous section.
So in many cases, results about three-dimensional $SU(2)$ gauge dynamics
are (indirectly) related to the phenomena under study here.

Case I:  If $h^{0}(V\vert_{C} \otimes {\cal O}(-1))$ and
$h^{0}(V_2 \vert_C \otimes {\cal O}(-1))$ are both greater than
two, then there is no worldsheet instanton generated superpotential
on the $\phi$ branch.  This follows from a standard counting of worldsheet
instanton zero modes \cite{Dist}, which are associated to these two
cohomology groups for the two kinds of noncritical strings.  
For these splitting types, there are too many zero modes for a potential
to be generated.  Furthermore, in these cases $\it fundamental$ string
worldsheet instanton effects also cannot remove the ``small instanton''
point in the classical moduli space, where the nontrivial fixed point
theory occurs.  The physics as analyzed at the scale $1/R$ persists
in the full quantum theory.  Such vacua provide examples where the
chirality change can occur while remaining in the moduli space of vacua,
i.e. at zero cost in energy. 

Case II:  If 
$h^{0}(V\vert_C \otimes {\cal O}(-1))$ or $h^{0}(V_2\vert_C \otimes
{\cal O}(-1))$ 
is $\it zero$, a nonperturbative worldsheet instanton superpotential
destabilizes the $\phi$ branch where the fivebrane is on the interval.
One can see this in the $E_8$ theory by considering tensionful string
worldsheet instantons of the first or second $E_8$.  The number of zero
modes on the worldsheet of such a string, for these splitting types, 
will be small enough for a nontrivial instanton-generated superpotential.
This is consistent with known results about $SU(2)$ gauge theories in
three dimensions \cite{ofer,ahw}.  
In this case, the chirality change which seems possible between the scale
$\mu$ generated by the instanton superpotential and $1/R$ is obstructed
by quantum effects below $\mu$.  However, the potential barrier which
prevents chirality change in such cases can be hierarchically smaller
than the string scale.  This indicates that in these cases chirality
change can ``almost'' occur, i.e. it is possible at a small cost in
energy.

In closing: The insights from string duality
have already been used to great effect in ``solving'' vacua with
extended supersymmetry.  In 4d N=1 
compactifications, there has been progress
in understanding methods of supersymmetry breaking
and in removing an obstruction to unifying vacua.
New techniques (particularly from F-theory/heterotic duality) are being
developed \cite{fmw,bpjs}, and 
one should expect further progress in the
near future.  

\centerline{\bf{Acknowledgements}}

Much of the work reported on in these notes was done in 
collaboration with E. Silverstein.  I would also like to thank
O. Aharony, P. Aspinwall, T. Banks, J. Distler, J. Louis, 
and N. Seiberg for discussions
on some of these topics.


\bigskip
\begin{thebibliography}{9}

\bibitem{revs} J. Polchinski, {\it Rev. Mod. Phys.} {\bf 68} (1996) 1245,
hep-th/9607050\semi
J. Schwarz, hep-th/9607201\semi
P. Aspinwall, hep-th/9611137\semi
C. Vafa, hep-th/9702201.
\bibitem{old} P. Green and T. Hubsch, {\it Phys. Rev. Lett.} {\bf 61}
(1988) 1163\semi
P. Candelas, P. Green, and T. Hubsch, {\it Phys. Rev. Lett.}
{\bf 62} (1989) 1956.
\bibitem{andy} A. Strominger, 
{\it Nucl. Phys.} {\bf B451} (1995) 96,
hep-th/9504090.
\bibitem{gms} B. Greene, D. Morrison, and A. Strominger,
{\it Nucl. Phys.} {\bf B451} (1995) 109, hep-th/9504145\semi
T. Chiang, B. Greene, M. Gross, and Y. Kanter, hep-th/9511204\semi
A. Avram, P. Candelas, D. Jancic, and M. Mandelberg, {\it Nucl. Phys.}
{\bf B465} (1996) 458, hep-th/9511230. 
\bibitem{kv} S. Kachru and C. Vafa, 
{\it Nucl. Phys.} {\bf B450}
(1995) 69, hep-th/9505105.
\bibitem{fhsv} S. Ferrara, J. Harvey, A. Strominger, and C. Vafa,
{\it Phys. Lett.} {\bf 361B} 
(1995) 59, hep-th/9505162.
\bibitem{chsw} P. Candelas, G. Horowitz, A. Strominger and E. Witten,
{\it Nucl. Phys.} {\bf B258} (1985) 46.
\bibitem{duy} K. Uhlenbeck and S.T. Yau, {\it Comm. Pure. App. Math.}
Vol XXXIX (1986) S257. 
\bibitem{hubsch} T. Hubsch, {\it Calabi-Yau Manifolds}, World Scientific, 
1992. 
\bibitem{dsww} M. Dine, N. Seiberg, X.G. Wen, and E. Witten,
{\it Nucl. Phys.} {\bf B278} (1986) 769, {\it Nucl. Phys.} {\bf B289}
(1987) 319. 
\bibitem{SW} E. Silverstein and E. Witten, {\it Nucl. Phys.} {\bf B444} (1995)
161, hep-th/9503212. 
\bibitem{typeI} J. Polchinski and E. Witten, {\it Nucl. Phys.} {\bf B460}
(1996) 525, hep-th/9510169. 
\bibitem{douglas} M. Douglas, hep-th/9512077. 
\bibitem{G2} See e.g. B.S. Acharya, {\it Nucl. Phys.} {\bf B475} (1996) 579,
hep-th/9603033 and references therein. 
\bibitem{VafaF} C. Vafa, {\it Nucl. Phys.} {\bf B469} (1996) 403, 
hep-th/9602022. 
\bibitem{Anom} C. Vafa and E. Witten, {\it Nucl. Phys.} {\bf B447} (1995)
261, hep-th/9505053. 
\bibitem{SVW} S. Sethi, C. Vafa and E. Witten, {\it Nucl. Phys.} {\bf B480}
(1996) 213, hep-th/9606122. 
\bibitem{Phases} E. Witten, {\it Nucl. Phys.} {\bf B403} (1993) 159, 
hep-th/9301042. 
\bibitem{DK} J. Distler and S. Kachru, {\it Nucl. Phys.} {\bf B413} (1994)
213, hep-th/9309110. 
\bibitem{DKtwo} J. Distler and S. Kachru, {\it Nucl. Phys.} {\bf B442} (1995)
64, hep-th/9501111. 
\bibitem{CDG} T. Chiang, J. Distler, and B. Greene, hep-th/9702030. 
\bibitem{CDGP} P. Candelas, X. De la Ossa, P. Green, and L. Parkes,
{\it Nucl. Phys.} {\bf B359} (1991) 21. 
\bibitem{Seiberg} N. Seiberg, {\it Phys. Rev.} {\bf D49} (1994) 6857,
hep-th/9402044. 
\bibitem{SeiWit} N. Seiberg and E. Witten, {\it Nucl. Phys.} {\bf B426}
(1994) 19, hep-th/9407087. 
\bibitem{chiral} S. Kachru and E. Silverstein, hep-th/9704185. 
\bibitem{bw} E. Witten and J. Bagger, {\it Phys. Lett.} 
{\bf 115B} (1982) 202. 
\bibitem{newsix} J. Blum and K. Intriligator, hep-th/9705044\semi
P. Aspinwall and D. Morrison, hep-th/9705104, and references therein.
\bibitem{small} E. Witten, {\it Nucl. Phys.} {\bf B460} (1996) 541, 
hep-th/9511030. 
\bibitem{ADHM} M. Atiyah, N. Hitchin, V. Drinfeld, and Y. Manin,
{\it Phys. Lett.} {\bf 65A} (1978) 185. 
\bibitem{hw} P. Horava and E. Witten, {\it Nucl. Phys.} {\bf B460} (1996) 506,
hep-th/9510209. 
\bibitem{GanHan} O. Ganor and A. Hanany, {\it Nucl. Phys.} {\bf B474} (1996) 
122, hep-th/9602120. 
\bibitem{SWstring} N. Seiberg and E. Witten, {\it Nucl. Phys.} {\bf B471}
(1996) 121, hep-th/9603003. 
\bibitem{dmw} M. Duff, R. Minasian, and E. Witten, {\it Nucl. Phys.} {\bf B465}
(1996) 413, hep-th/9601036. 
\bibitem{seibetal} N. Seiberg, {\it Phys. Lett.} {\bf B388} (1996) 753,
hep-th/9608111\semi
D. Morrison and N. Seiberg, {\it Nucl. Phys.} {\bf B483} (1997) 229,
hep-th/9609070\semi
M. Douglas, S. Katz, and C. Vafa, hep-th/9609071\semi
O. Ganor, D. Morrison and N. Seiberg, {\it Nucl. Phys.} {\bf B487} (1997)
93, hep-th/9610251. 
\bibitem{kss} S. Kachru, N. Seiberg, and E. Silverstein, {\it Nucl. Phys.}
{\bf B480} (1996) 170, hep-th/9605036. 
\bibitem{ads} I. Affleck, M. Dine, and N. Seiberg, {\it Nucl. Phys.} {\bf
B241} (1984) 493. 
\bibitem{Dist} J. Distler, {\it Phys. Lett.} {\bf B188} (1987) 431. 
\bibitem{ks} S. Kachru and E. Silverstein, {\it Nucl. Phys.} {\bf B482}
(1996) 92, hep-th/9608194. 
\bibitem{witsup} E. Witten, {\it Nucl. Phys.} {\bf B474} (1996) 343,
hep-th/9604030. 
\bibitem{dgw} R. Donagi, A. Grassi, and E. Witten, {\it Mod. Phys.
Lett.} {\bf A11} (1996) 2199, hep-th/9607091.
\bibitem{cl} G. Curio and D. Lust, hep-th/9703007. 
\bibitem{witwrap} E. Witten, {\it Nucl. Phys.} {\bf B471} (1996) 135,
hep-th/9602070. 
\bibitem{aks} O. Aharony, S. Kachru, and E. Silverstein,
{\it Nucl. Phys.} {\bf B488} (1997) 159, hep-th/9610205.
\bibitem{ofer} O. Aharony, A. Hanany, K. Intriligator, N. Seiberg and
M. Strassler, hep-th/9703110. 
\bibitem{ahw} I. Affleck,  J. Harvey and E.  Witten, {\it Nucl. Phys.}
{\bf B206} (1982) 413. 
\bibitem{fmw} R. Friedman, J. Morgan, and E. Witten, hep-th/9701162.
\bibitem{bpjs} M. Bershadsky, A. Johansen, T. Pantev, and V. Sadov,
hep-th/9701165. 

\end{thebibliography}
\end{document}